\documentclass[a4paper,10pt]{article}
\usepackage{amsfonts}
\usepackage{amsmath}
\usepackage{amssymb}
\usepackage{graphicx}

\begin{document}

\title{Statistical repulsion/attraction of electrons in graphene in a
magnetic field}
\author{J.S. Ardenghi$^{\dag }$\thanks{%
email:\ jsardenghi@gmail.com, fax number:\ +54-291-4595142}, P. Bechthold$%
^{\dag }$, E. Gonzalez$^{\dag }$, P. Jasen$^{\dag }$ and A. Juan$^{\dag }$ \\
$^{\dag }$IFISUR, Departamento de F\'{\i}sica (UNS-CONICET)\\
Avenida Alem 1253, Bah\'{\i}a Blanca, Argentina}
\maketitle

\begin{abstract}
The aim of this work is to describe the thermodynamic properties of an
electron gas in graphene placed in a constant magnetic field. The electron
gas is constituted by $N$ Bloch electrons in the long wavelength
approximation. The partition function is analyzed in terms of a perturbation
expansion of the dimensionless constant $(\sqrt{eB}L)^{-1}$. The statistical
repulsion/attraction potential for electrons in graphene is obtained in the
respective case in which antisymmetric/symmetric states in the coordinates
are chosen. Thermodynamic functions are computed for different orders in the
perturbation expansion and the different contributions are compared for
symmetric and antisymmetric states, showing remarkable differences between
them due to the spin exchange correlation. A detailed analysis of the
statistical potential is done, showing that, although electrons satisfy
Fermi statistics, attractive potential at some interparticle distances can
be found.
\end{abstract}

\section{Introduction}

Graphene is a two-dimensional allotrope of carbon which has become one of
the most significant topics in solid state physics due to the large number
of applications (\cite{novo},\cite{intro1},\cite{intro2}, \cite{B}, \cite%
{BBBB}). The carbon atoms form a honey-comb lattice made of two
interpenetrating triangular sublattices, $A$ and $B$. A special feature of
the graphene band structure is the linear dispersion at the Dirac points
which are dictated by the $\pi $ and $\pi ^{\prime }$ bands that form
conical valleys touching at the high symmetry points of the Brillouin zone 
\cite{A}. Electrons near these symmetry points behave as massless
relativistic Dirac fermions with an effective Dirac-Weyl Hamiltonian \cite{B}%
. When a magnetic field is applied perpendicular to the graphene sheet, a
discretization of the energy levels is obtained, the so called Landau levels 
\cite{kuru}. These quantized energy levels still appear also for
relativistic electrons, just their dependence on field and quantization
parameter is different. In a conventional non-relativistic electron gas,
Landau quantization produces equidistant energy levels, which is due to the
parabolic dispersion law of free electrons. In graphene, the electrons have
relativistic dispersion law, which strongly modifies the Landau quantization
of the energy and the position of the levels. In particular, these levels
are not equidistant as occurs in a conventional non-relativistic electron
gas in a magnetic field. This large gap allows one to observe the quantum
Hall effect in graphene, even at room temperature \cite{E}.

The thermodynamics properties of graphene and graphene nanoribbons have been
studied under electric and magnetic modulations from the theoretical and
experimental viewpoint (see \cite{NA1}, \cite{NA2}, \cite{NA3}, \cite{NA4}
and \cite{NA6}) The presence of the electric and magnetic modulation expands
the Landau energy levels into bands and these bandwidths oscillates with the
electric and magnetic fields (Weiss oscillation, \cite{NA5}). Also, magnetic
oscillation can be present in the zigzag ribbons. At large width, the low
field oscillations for zigzag ribbons is much faster that that of armchair
ribbons. In turn, in doped gapped graphene the electronic heat capacity
shows the Schottky anomaly typical for low temperatures systems \cite{hamze}%
. Following the line of these previous works, this paper is concerned with
the exact description of the quantum partition function of a collection of $%
N $ Bloch electrons in graphene, in the long wavelength approximation,
placed in a constant magnetic field. The classical limit and the lowest
quantum corrections are computed. We will define a complete $N$-body wave
function with the antisymmetrized/symmetrized product of a set of single
particle wave functions, which are the eigenfunctions of Bloch electrons in
the long wavelength approximation placed in a magnetic field. In this sense,
the procedure to define the partition function and the quantum corrections
will be identical to the procedure which appears in textbooks (for example 
\cite{Huang}), with the difference that the set of single particle wave
functions used in these textbooks are plane waves. Finally, the entropy,
internal energy, specific heat and magnetization can be computed for
different orders of the partition function.

For a self-contained lecture of this paper, a brief introduction of the
quantum mechanics of graphene in a constant magnetic field in the long
wavelength approximation can be introduced (see \cite{B}). The Hamiltonian
in the two inequivalent corners of the Brillouin zones reads%
\begin{equation}
H=v_{f}\left( 
\begin{array}{cccc}
0 & p_{x}-ip_{y} & 0 & 0 \\ 
p_{x}+ip_{y} & 0 & 0 & 0 \\ 
0 & 0 & 0 & -p_{x}-ip_{y} \\ 
0 & 0 & -p_{x}+ip_{y} & 0%
\end{array}%
\right)  \label{intro1}
\end{equation}%
where $\overrightarrow{p}=\overrightarrow{k}-e\overrightarrow{A}$ is the
quasiparticle momentum, $e$ is the electron charge, $\overrightarrow{A}$ is
the vector potential which in the Landau gauge reads $\overrightarrow{A}%
=(-By,0,0)$ and $v_{f}=10^{6}m/s$ is the Fermi velocity (in this work we
will use $c=\hbar =1$). The eigenfunctions and eigenvectors for the
Hamiltonian of last equation reads%
\begin{equation}
\psi _{(n,s,k)}(r)=e^{ikx}\frac{C_{n}}{\sqrt{2L}}\varphi _{(n,s,k)}(\xi )
\label{intro2}
\end{equation}%
where $\varphi _{(n,s,k)}(\xi )$ reads%
\begin{equation}
\varphi _{(n,s,k)}(\xi )=\left( 
\begin{array}{c}
-s\phi _{n-1,k}(\xi )(1-\delta _{n,0}) \\ 
\phi _{n,k}(\xi ) \\ 
\phi _{n,k}(\xi ) \\ 
s\phi _{n-1,k}(\xi )(1-\delta _{n,0})%
\end{array}%
\right)  \label{intro3}
\end{equation}%
being $\phi _{n,k}(\xi )$ the wave function of the harmonic oscillator%
\footnote{%
The factor $(1-\delta _{n,0})\ $is introduced to discriminate the wave
function with $n=0$. In this case, only one sublattice contributes in both
valleys $K$ and $K^{\prime }$.}%
\begin{equation}
\phi _{n,k}(\xi )=\frac{\pi ^{-1/4}}{\sqrt{2^{n}n!}}e^{-\frac{1}{2}\xi
^{2}}H_{n_{,k}}(\xi )  \label{intro4}
\end{equation}%
and $\xi =\frac{y}{l_{B}}-l_{B}k$, $L=\sqrt{A}$ where $A~$is the area of the
graphene sheet and $s=\pm 1$ is the conduction (valence) band index. The
coefficient $C_{n}$ is $C_{n}=\frac{1}{\sqrt{2-\delta _{n,0}}}$ and $l_{B}=%
\sqrt{1/eB}$ is the magnetic length. The eigenvalues of the Hamiltonian reads%
\begin{equation}
E_{n,s}=s\Omega \sqrt{n}  \label{intro5}
\end{equation}%
where $\Omega =\sqrt{2}v_{f}/l_{B}$. The low energy description is only
valid as long as the characteristic energy of the excitations is not larger
than an energy cutoff $E_{n,1}<\Delta $, where $\ \Delta =v_{f}k_{\Delta }$
and $k_{\Delta }~$is a momentum cutoff. A simple way to choose $k_{\Delta }$
is by the condition imposed by the linear term in the Taylor expansion of
the energy, that is, $k_{\Delta }<\frac{1}{a}$ where $a$ is the lattice
spacing. Another slightly different, but more exact way, is by choosing $%
k_{\Delta }$ in such a way to conserve the total number of states in the
Brillouin zone, that is, $\pi k_{\Delta }^{2}=(2\pi )^{2}/A_{C}$, where $%
A_{C}=3\sqrt{3}a^{2}/2$ is thea area of the hexagonal lattice (see \cite%
{peres-guinea}). Then, using eq.(\ref{intro5}), $E<\Delta $ implies that $%
n=n_{\Delta }<\frac{4\pi }{3\sqrt{3}a^{2}eB}$, then for weak magnetic
fields, the cutoff tends to infinity and for high magnetic fields, the
cutoff tends to zero.

With the eigenfunctions and eigenvalues of an electron in graphene we can
apply the machinery of statistical mechanics by computing the partition
function under succesive permutations.

The work will be organized as follows:

In Section 2, the partition function for $N$ Bloch electrons will be
introduced and computed using the results found in Appendix A. A detailed
description of the succesive terms of the perturbation expansion is done up
to two permutations. Exact results are found for $p=0$ and $p=1$
permutations and for $p=2$ an integral equation is obtained

In Section 3, a comparation for the entropy, internal energy, specific heat
and magnetization is shown for $p=0$ and $p=1$ for antisymmetric and
symmetric states in the coordinates. The differences between these
thermodynamic functions in terms of temperature are analyzed, showing how
the exchange correlation introduce unexpected features in the internal
energy and entropy of the quantum system.

In section 4, the conclusions are presented.

In Appendix A, a detailed description of the functions involved in the
partition function are computed.

\section{Partition function for $N$ Bloch electrons}

The partition function of $N$ Bloch electrons in the long wavelength
approximation in graphene sheet under a constant magnetic field reads%
\begin{equation}
Z_{N}(\beta
)=\sum\limits_{s_{1}=-1}^{+1}...\sum\limits_{s_{N}=-1}^{+1}\sum%
\limits_{n_{1}=0}^{+n_{\Delta }}...\sum\limits_{n_{N}=0}^{+n_{\Delta }}\exp
(-\beta \sum\limits_{j=1}^{N}E_{n_{j},s_{j}})\times \int \chi _{\alpha
_{1},...,\alpha _{N}}(r_{1},...,r_{N})d^{2}r_{1}...d^{2}r_{N}  \label{new1}
\end{equation}%
where $\chi _{\alpha _{1},...,\alpha _{N}}(r_{1},...,r_{N})$ is a function
of the position of electrons that reads%
\begin{equation}
\chi _{\alpha _{1},...,\alpha _{N}}(r_{1},...,r_{N})=\frac{1}{N!}%
\sum\limits_{P}^{{}}\sum\limits_{P^{\prime }}^{{}}\delta _{P}\delta
_{P^{\prime }}\overset{N}{\underset{j=1}{\prod }}f_{\alpha
_{j}}(Pr_{j},P^{\prime }r_{j})  \label{new2}
\end{equation}%
where\footnote{%
See Appendix A for the deduction of formulas of this section.} 
\begin{equation}
f_{\alpha _{j}}(Pr_{j},P^{\prime }r_{j})=\int dk_{j}\psi _{\alpha
_{j}}^{\ast }(Pr_{j})\psi _{\alpha _{j}}(P^{\prime }r_{j})=-\frac{e^{i\sigma
(Pr_{j},P^{\prime }r_{j})}}{(2-\delta _{n,0})Ll_{B}}g_{n_{j}}(\left\vert
Pr_{j}-P^{\prime }r_{j}\right\vert ^{2})  \label{new3}
\end{equation}%
and where $\sigma $ is a gauge-dependent function that reads%
\begin{equation}
\sigma (Pr_{j},P^{\prime }r_{j})=\frac{1}{2l_{B}^{2}}((P^{\prime
}y_{j}+Py_{j})(P^{\prime }x_{j}-Px_{j}))  \label{new4}
\end{equation}%
and%
\begin{gather}
g_{n_{j}}(\left\vert Pr_{j}-P^{\prime }r_{j}\right\vert ^{2})=e^{-\frac{1}{%
4l_{B}^{2}}\left\vert Pr_{j}-P^{\prime }r_{j}\right\vert ^{2}}\times
\label{new5} \\
\left[ L_{n_{j}-1}(\frac{1}{2l_{B}^{2}}\left\vert Pr_{j}-P^{\prime
}r_{j}\right\vert ^{2})(1-\delta _{n_{j},0})+L_{n_{j}}(\frac{1}{2l_{B}^{2}}%
\left\vert Pr_{j}-P^{\prime }r_{j}\right\vert ^{2})\right]  \notag
\end{gather}%
where $L_{n}(x)$ is the Laguerre polynomial of order $n$ (see Appendix A).
The result found in las equation are similar of those found in \cite{dyna1}
and \cite{dyna2} for the Green function.

If we change the permutation index $P$ by $P^{\prime }$, the function $%
f_{\alpha _{j}}$ only changes in the sign of the exponential $e^{i\sigma
(Pr_{j},P^{\prime }r_{j})}$, that is, the gauge function is antisymmetric
under the interchange of its coordinates 
\begin{equation}
\sigma (Pr_{j},P^{\prime }r_{j})=-\sigma (P^{\prime }r_{j},Pr_{j})
\label{bla3.1.2}
\end{equation}%
then the sum in the permutation $P^{\prime }$ in eq.(\ref{new2}) gives the
same contribution as the sum in $P$ with a minus sign in $\sigma $, that is%
\begin{gather}
f_{\alpha _{j}}(Pr_{j},P^{\prime }r_{j})+f_{\alpha _{j}}(P^{\prime
}r_{j},Pr_{j})=-\frac{1}{(2-\delta _{n_{j},0})Ll_{B}}g_{n_{j}}(\left\vert
Pr_{j}-P^{\prime }r_{j}\right\vert ^{2})\left( e^{i\sigma (Pr_{j},P^{\prime
}r_{j})}+e^{-i\sigma (Pr_{j},P^{\prime }r_{j})}\right)  \label{bla3.1.3} \\
=-\frac{2}{(2-\delta _{n_{j},0})Ll_{B}}g_{n_{j}}(\left\vert Pr_{j}-P^{\prime
}r_{j}\right\vert ^{2})\cos (\sigma (Pr_{j},P^{\prime }r_{j}))  \notag
\end{gather}%
Taking this into account, the function $\chi _{\alpha _{1},...,\alpha
_{N}}(r_{1},...,r_{N})$ can be written in a more useful form as%
\begin{equation}
\chi _{\alpha _{1},...,\alpha _{N}}(r_{1},...,r_{N})=\frac{(-1)^{N}}{N!}%
\frac{2^{N}}{(Ll_{B})^{N}}\sum\limits_{P}^{{}}\delta _{P}\overset{N}{%
\underset{j=1}{\prod }}\frac{g_{n_{j}}(\left\vert Pr_{j}-r_{j}\right\vert
^{2})}{(2-\delta _{n_{j},0})}\cos (\sum\limits_{i=1}^{N}\sigma
(Pr_{i},r_{i}))  \label{bla3.1.4}
\end{equation}%
In textbooks, a thermodynamics limit is taken (see \cite{Huang}, pag. 202),
where the interparticle distance is much larger than the thermal wavelength.
In this case, the temperature does not appear in eq.(\ref{bla3.1.4}) and a
thermodynamics limit cannot be taken. Nevertheless, a different
approximation can be applied, where the graphene sheet area is much larger
than the area defined by the magnetic length $l_{B}$, that is, $%
L^{2}>>>l_{B}^{2}$, then the interparticle distance $\left\vert
r_{i}-r_{j}\right\vert $ can be larger than $l_{B}$, this is $%
l_{B}<\left\vert r_{i}-r_{j}\right\vert <L$\thinspace . The sum in $P$ in
last equation contains $N!$ terms that can be arranged as a sum with
increased number of permutations, then we can write 
\begin{equation}
\chi _{\alpha _{1},...,\alpha _{N}}(r_{1},...,r_{N})=\eta
_{N}\sum\limits_{p=0}^{N-1}(-1)^{p}\gamma
_{n_{1},...,n_{N}}^{(p)}(r_{1},...,r_{N})  \label{bla3.1.7}
\end{equation}%
where $\eta _{N}~$is a constant with units of area$^{-N}$ which reads%
\begin{equation}
\eta _{N}=\frac{(-1)^{N}}{N!}\frac{2^{N}}{(Ll_{B})^{N}}  \label{bla3.1.7.1}
\end{equation}%
and $p$ is an index that counts the number of permutations. The factor $%
(-1)^{p}$ contains the sign of odd and even permutations. In the case that
the antisymmetric state is contained in the spin variables and not over the
coordinates, the factor $(-1)^{p}$ do not appears in eq.(\ref{bla3.1.7}). In
particular, the term without permutation reads

\begin{equation}
\gamma _{n_{1},...,n_{N}}^{(0)}(r_{1},...,r_{N})=2^{N}  \label{bla3.1.7.1.1}
\end{equation}%
and with one permutation\footnote{%
These results will be obtained in the next sections.}%
\begin{equation}
\gamma
_{n_{1},...,n_{N}}^{(1)}(r_{1},...,r_{N})=2^{N-2}\sum\limits_{i=1}^{N}\sum%
\limits_{j=i+1}^{N}g_{n_{i}}(\left\vert r_{j}-r_{i}\right\vert
^{2})g_{n_{j}}(\left\vert r_{i}-r_{j}\right\vert ^{2})  \label{bla3.1.7.2}
\end{equation}%
These two last results\ and the limit $\left\vert r_{i}-r_{j}\right\vert
>l_{B}$ can be used to apply the following approximation%
\begin{gather}
2^{N}(1\pm \frac{1}{4}\gamma _{n_{1},...,n_{N}}^{(1)}(r_{1},...,r_{N}))\sim
2^{N}\overset{N}{\underset{i=1}{\prod }}\overset{N}{\underset{j=i+1}{\prod }}%
(1\pm \frac{1}{4}g_{n_{i}}(\left\vert r_{j}-r_{i}\right\vert
^{2})g_{n_{j}}(\left\vert r_{i}-r_{j}\right\vert ^{2}))=  \label{bla1.3.7.3}
\\
\exp (-\beta \sum\limits_{i=1}^{N}\sum\limits_{j\neq
i}^{N}V_{n_{i},n_{j}}(r_{i},r_{j}))  \notag
\end{gather}%
where $V_{n_{i},n_{j}}(r_{i},r_{j})$ is the statistical repulsion/attraction
interparticle potential which reads%
\begin{equation}
V_{n_{i},n_{j}}(r_{i},r_{j})=-\frac{1}{\beta }\ln (1\pm \frac{1}{4}%
g_{n_{i}}(\left\vert r_{j}-r_{i}\right\vert ^{2})g_{n_{j}}(\left\vert
r_{i}-r_{j}\right\vert ^{2}))  \label{bla1.3.8}
\end{equation}%
where the plus (minus) sign is for symmetric (antisymmetric) states in the
coordinates. These two results give the statistical repulsion/attraction for
fermions in graphene in a constant magnetic field in the limit $\left\vert
r_{i}-r_{j}\right\vert >l_{B}$. Actually, the interparticle potential will
vary with the permutation order and cannot be disentagled into two-particle
potential. 
\begin{figure}[tbp]
\centering
\includegraphics[width=105mm,height=70mm]{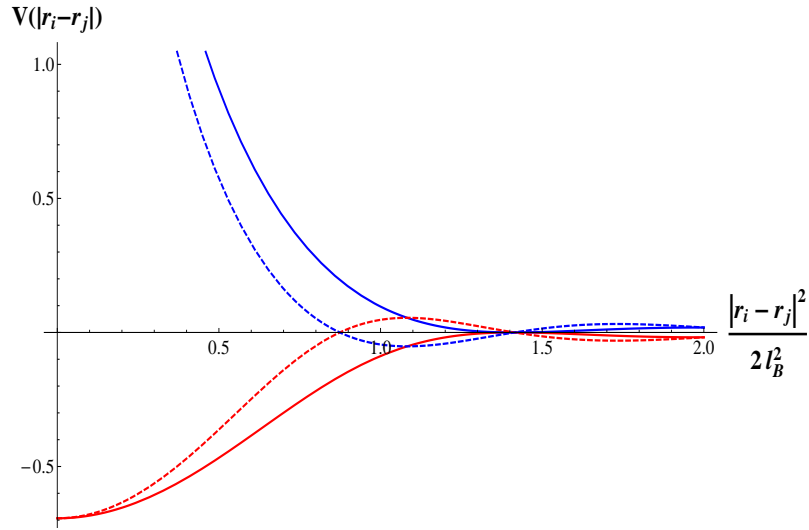}
\caption{Statistical attraction potential (red line for symmetric states in
the coordinates) and statistical repulsion potential (blue line for
antisymmetric states in the coordinates) for electrons in graphene. $n_{i}=1$
and $n_{j}=1$ for continuous line and $n_{i}=2$ and $n_{j}=1$ for dashed
line.}
\label{figura0}
\end{figure}
In the case of weak magnetic fields, the cutoff tends to $n_{\Delta
}\rightarrow \infty $, then we can take the limit of large quantum numbers
of the statistical potential (see \cite{laguerre}, page 1003)%
\begin{equation}
\underset{n_{i},n_{j}\rightarrow \infty }{\lim }V_{n_{i},n_{j}}(r_{i,}r_{j})%
\sim -\frac{1}{\beta }\ln (1\pm 4e^{-\frac{1}{2l_{B}^{2}}\left\vert
r_{i}-r_{j}\right\vert ^{2}}\frac{\sqrt{2}l_{B}}{\left\vert
r_{i}-r_{j}\right\vert })  \label{bla1.3.8.1}
\end{equation}%
which differs from the interparticle statistical potential of eq.(9.57) of 
\cite{Huang} for the Coulomb factor. An interesting result is that for
antisymmetric states in the coordinates, the logarithm function diverges
when the argument is zero, which gives the following equation for the
minimum distance for the repulsive potential between electrons in the limit
of weak magnetic fields and large quantum numbers%
\begin{equation}
\ln \frac{x}{4}=-x^{2}  \label{bla1.3.8.2}
\end{equation}%
where $x=\frac{\left\vert r_{i}-r_{j}\right\vert }{2l_{B}}$. The solution of
last equation is $x=\sqrt{W(2)/2}\sim 0.652$ where $W(x)$ is the Lambert $W$%
-function (see \cite{lambert}), then $\left\vert r_{i}-r_{j}\right\vert
>1.304l_{B}$.

As a final consideration for this section, we can introduce eq.(\ref%
{bla3.1.7}) in eq.(\ref{new1}) and because the argument inside the integral
do not depends on $s_{j}$, then the summation on this label can be done and
the result reads 
\begin{equation}
Z_{N}(\beta )=\sum\limits_{n_{1}=0}^{+n_{\Delta
}}...\sum\limits_{n_{N}=0}^{+n_{\Delta }}\eta
_{N}A_{n_{1},...,n_{N}}(N,\beta ,\Omega )B_{n_{1},...,n_{N}}(N)
\label{bla1.5}
\end{equation}%
where%
\begin{equation}
A_{n_{1},...,n_{N}}(N,\beta ,\Omega )=2^{N}\overset{N}{\underset{j=1}{\prod }%
}\mu _{n}\cosh (\beta \Omega \sqrt{n_{j}})  \label{bla1.6}
\end{equation}%
where%
\begin{equation}
\mu _{n}=(2-\delta _{n,0})^{-1}  \label{bla1.6.1}
\end{equation}%
and%
\begin{equation}
B_{n_{1},...,n_{N}}(N)=\sum\limits_{p=0}^{N-1}\int (-1)^{p}\gamma
_{n_{1},...,n_{N}}^{(p)}(r_{1},...,r_{N})d^{2}r_{1}...d^{2}r_{N}
\label{bla1.7}
\end{equation}%
In next sections, the first three terms of last equation will be obtained
and general considerations will be done for the remaining terms.

\subsection{$p=0$ permutation}

If we consider no permutation in the coordinates, then the partition
function reads%
\begin{equation}
Z_{N}^{(0)}(\beta )=\sum\limits_{n_{1}=0}^{+n_{\Delta
}}...\sum\limits_{n_{N}=0}^{+n_{\Delta }}2^{N}\overset{N}{\underset{j=1}{%
\prod }}\mu _{n}\cosh (\beta \Omega \sqrt{n_{j}})\eta _{N}L^{2N}2^{N}
\label{bla1.3.9}
\end{equation}%
where we have integrated in $d^{2}r_{j}$ and $L^{2}$ is the area of the
graphene sheet. In this case, the function $\gamma
_{n_{1},...,n_{N}}^{(0)}(r_{1},...,r_{N})$ reads%
\begin{equation}
\gamma _{n_{1},...,n_{N}}^{(0)}(r_{1},...,r_{N})=L^{2N}2^{N}
\label{bla1.3.9.0.0}
\end{equation}%
We can separate each sum in $n_{j}$ and because they are equal we obtain 
\begin{equation}
Z_{N}^{(0)}(\beta )=\eta _{N}2^{2N}L^{2N}\left(
\sum\limits_{n=0}^{+n_{\Delta }}\mu _{n}\cosh \left( \beta \Omega \sqrt{n}%
\right) \right) ^{N}  \label{bla1.3.9.1}
\end{equation}%
which is quite similar to the partition function of a magnetic system. From
this partition function we can compute the entropy, the internal energy, the
heat capacity and magnetization by using the Helmholtz free energy $F=-kT\ln
Z_{N}$ by the following equations%
\begin{eqnarray}
S &=&-\frac{\partial F}{\partial T}\text{\ \ \ \ \ \ \ \ \ }M=-\frac{%
\partial F}{\partial B}  \label{bla1.3.9.2} \\
U &=&F+TS\text{ \ \ \ \ \ \ }C=\frac{\partial U}{\partial T}  \notag
\end{eqnarray}%
In turn, using the Helmholtz free energy we can obtain the pressure of the
thermodynamic system by the following equation $P=-\frac{\partial F}{%
\partial A}$, where $A=L^{2}$.\ Using the partition function obtained in eq.(%
\ref{bla1.3.9.1}), the pressure for a gas of Bloch electrons in the
classical limit of the partition function reads%
\begin{equation}
PA=\frac{N}{2}kT  \label{bla1.3.9.4}
\end{equation}%
which is almost identical to state equation for and ideal gas in two
dimension.

\subsection{One permutation $p=1$}

For one permutation, we have to take into account the function $\gamma
_{n_{1},...,n_{N}}^{(1)}(r_{1},...,r_{N})$ which can be integrated in the
coordinates 
\begin{equation}
\int \gamma
_{n_{1},...,n_{N}}^{(1)}(r_{1},...,r_{N})d^{2}r_{1}...d^{2}r_{N}=L^{2(N-2)}2^{N-2}\sum\limits_{i=1}^{N}\sum\limits_{j=i+1}^{N}Q_{n_{i},n_{j}}
\label{one1}
\end{equation}%
where%
\begin{equation}
Q_{n_{i},n_{j}}=\int g_{n_{i}}(\left\vert r_{j}-r_{i}\right\vert
^{2})g_{n_{j}}(\left\vert r_{i}-r_{j}\right\vert ^{2})d^{2}r_{i}d^{2}r_{j}
\label{one2}
\end{equation}%
The factor $L^{2(N-2)}2^{N-2}$ appears due to the integration of the
non-permutted coordinates. The function $Q_{n_{i},n_{j}}$ of similar to
eq.(8) of \cite{tapash}, which is used as an interaction potential for
electrons within a single Landau level. This potential is determined by the
relative strength of the Coulomb interaction within the $n$-Landau level and
it is used to study the charge excitacions of quantum ferromagnetic states
(see \cite{nomura} and \cite{goerbig2}). In fact, the integral of eq.(\ref%
{bla3.1.4}) is the exact interaction potential between $N$ electrons at
different Landau levels.Using the result of eq.(\ref{bla3.1.4}) we can write 
$Q_{n_{i},n_{j}}$ as%
\begin{equation}
Q_{n_{i},n_{j}}=\sum\limits_{l=0}^{1}\sum%
\limits_{k=0}^{1}W_{n_{i}-l,n_{j}-k}(1-\delta _{n_{i},0})^{l}(1-\delta
_{n_{j},0})^{k}  \label{one3}
\end{equation}%
where $W_{a,b}$ reads%
\begin{equation}
W_{a,b}=\int e^{-\frac{1}{2l_{B}^{2}}\left\vert r_{i}-r_{j}\right\vert
^{2}}L_{a}(\frac{1}{2l_{B}^{2}}\left\vert r_{i}-r_{j}\right\vert ^{2})L_{b}(%
\frac{1}{2l_{B}^{2}}\left\vert r_{i}-r_{j}\right\vert
^{2})d^{2}r_{i}d^{2}r_{j}  \label{one4}
\end{equation}%
writing $\left\vert r_{i}-r_{j}\right\vert
^{2}=(x_{i}-x_{j})^{2}+(y_{i}-y_{j})^{2}$ and making the following center of
mass coordinate transformation 
\begin{eqnarray}
x_{i}-x_{j} &=&x_{ij}\text{ \ \ \ \ \ \ \ \ \ }X_{ij}=\frac{1}{2}%
(x_{i}+x_{j})  \label{one5} \\
y_{i}-y_{j} &=&y_{ij}\text{ \ \ \ \ \ \ \ \ \ }Y_{ij}=\frac{1}{2}%
(y_{i}+y_{j})  \notag
\end{eqnarray}%
then%
\begin{equation}
W_{a,b}=L^{2}\int e^{-\frac{1}{2l_{B}^{2}}(x_{ij}^{2}+y_{ij}^{2})}L_{a}(%
\frac{1}{2l_{B}^{2}}(x_{ij}^{2}+y_{ij}^{2}))L_{b}(\frac{1}{2l_{B}^{2}}%
(x_{ij}^{2}+y_{ij}^{2}))dx_{ij}dy_{ij}  \label{one6}
\end{equation}%
where the area $L^{2}$ appears due to the $dX_{ij}$ and $dY_{ij}$
integration. Finally, using polar coordinates $x_{ij}=r_{ij}\cos \theta
_{ij} $ and $y_{ij}=r_{ij}\sin \theta _{ij}$, and performing the $\theta
_{ij}$ integration we obtain%
\begin{equation}
W_{a,b}=2\pi L^{2}\int e^{-\frac{r_{ij}^{2}}{2l_{B}^{2}}}L_{a}(\frac{%
r_{ij}^{2}}{2l_{B}^{2}})L_{b}(\frac{r_{ij}^{2}}{2l_{B}^{2}})r_{ij}dr_{ij}
\label{one7}
\end{equation}%
finally, making the coordinate transformation $s=r_{ij}^{2}/2l_{B}^{2}$ the
function $W_{a,b}$ reads%
\begin{equation}
W_{a,b}=2\pi (Ll_{B})^{2}\int e^{-s}L_{a}(s)L_{b}(s)ds  \label{one8}
\end{equation}%
using the orthogonality of the Laguerre polynomials we obtain%
\begin{equation}
W_{a,b}=2\pi (Ll_{B})^{2}\int e^{-s}L_{a}(s)L_{b}(s)ds=2\pi
(Ll_{B})^{2}\delta _{ab}  \label{one9}
\end{equation}%
Replacing this last result in eq.(\ref{one2}) we obtain%
\begin{equation}
Q_{n_{i},n_{j}}=\sum\limits_{l=0}^{1}\sum%
\limits_{k=0}^{1}W_{n_{i}-l,n_{j}-k}(1-\delta _{n_{i},0})^{l}(1-\delta
_{n_{j},0})^{k}=2\pi
(Ll_{B})^{2}\sum\limits_{l=0}^{1}\sum\limits_{k=0}^{1}\delta
_{n_{i}-l,n_{j}-k}(1-\delta _{n_{i},0})^{l}(1-\delta _{n_{j},0})^{k}
\label{one10}
\end{equation}%
Then the partition function with the correction given by one permutation
reads

\begin{gather}
Z_{N}^{(1)}(\beta )=\sum\limits_{n_{1}=0}^{+n_{\Delta
}}...\sum\limits_{n_{N}=0}^{+n_{\Delta }}2^{N}\overset{N}{\underset{j=1}{%
\prod }}\mu _{n}\cosh (\beta \Omega \sqrt{n_{j}})\eta _{N}\times
\label{one10.1} \\
\left( L^{2N}2^{N}+\frac{\pi }{2}L^{2(N-1)}2^{N-2}l_{B}^{2}\sum%
\limits_{i=1}^{N}\sum\limits_{j=i+1}^{N}\sum\limits_{l=0}^{1}\sum%
\limits_{k=0}^{1}\delta _{n_{i}-l,n_{j}-k}(1-\delta _{n_{i},0})^{l}(1-\delta
_{n_{j},0})^{k}\right)  \notag
\end{gather}%
By doing some complex algebraic manipulations is possible to obtain a exact
description of the contribution of one permutation to the partition
function, which reads

\begin{equation}
Z_{N}^{(1)}(\beta )=Z_{N}^{(0)}(\beta )\left( 1+\frac{\pi }{8}\frac{l_{B}^{2}%
}{L^{2}}F_{1}(\beta ,\Omega ,N)\right)  \label{one10.1.1}
\end{equation}%
where%
\begin{equation}
F_{1}(\beta ,\Omega ,N)=N(N-1)\frac{\sum\limits_{n=0}^{+n_{\Delta }}\mu
_{n}\cosh ^{2}(\beta \Omega \sqrt{n})+\sum\limits_{n=1}^{+n_{\Delta }}\mu
_{n}\mu _{n-1}\cosh (\beta \Omega \sqrt{n})\cosh (\beta \Omega \sqrt{n-1}%
)-1/2}{\left( \sum\limits_{n=0}^{+n_{\Delta }}\mu _{n}\cosh (\beta \Omega 
\sqrt{n})\right) ^{2}}  \label{one1.1.2}
\end{equation}%
The correction introduced by eq.(\ref{one1.1.2}) can be used to obtain the
thermodynamic functions and to compare it with the results obtained in last
subsection.

\subsection{Two permutation $p=2$}

The contribution to the partition function of two permutations is computed.
In this case, an interesting effect appears due to the gauge dependent term $%
\sigma $ when we take into account two permutations%
\begin{gather}
\int \gamma
_{n_{1},...,n_{N}}^{(2)}(r_{1},...,r_{N})d^{2}r_{1}...d^{2}r_{N}=L^{2(N-3)}2^{N-3}\times
\label{two0} \\
\sum\limits_{i=1}^{N}\sum\limits_{j=i+1}^{N}\sum%
\limits_{k=j+1}^{N}U_{n_{i},n_{j},n_{k}}  \notag
\end{gather}%
where%
\begin{equation}
U_{n_{i},n_{j},n_{k}}=\int g_{n_{i}}(\left\vert r_{j}-r_{i}\right\vert
^{2})g_{n_{j}}(\left\vert r_{k}-r_{j}\right\vert ^{2})g_{n_{k}}(\left\vert
r_{i}-r_{k}\right\vert ^{2})\cos (\sigma (r_{j},r_{i})+\sigma
(r_{k},r_{j})+\sigma (r_{i},r_{k})))d^{2}r_{i}d^{2}r_{j}d^{2}r_{k}
\label{two1}
\end{equation}%
Last equation prevents to consider the statistical potential of eq.(\ref%
{bla1.3.8}) beyond the first order in the perturbation expansion due to the
gauge dependent term~$\sigma $ that appears when we consider at least three
particles. Using eq.(\ref{new4}), the argument of the cosine function reads%
\begin{equation}
\sigma (r_{j},r_{i})+\sigma (r_{k},r_{j})+\sigma (r_{i},r_{k})=\frac{1}{%
2l_{B}^{2}}\left(
y_{j}x_{i}-y_{i}x_{j}+y_{k}x_{j}-y_{j}x_{k}+y_{i}x_{k}-y_{k}x_{i}\right)
\label{two2}
\end{equation}%
which can be rewritten in terms of skew products between position vectors%
\begin{equation}
\sigma (r_{j},r_{i})+\sigma (r_{k},r_{j})+\sigma (r_{i},r_{k})=\frac{1}{%
2l_{B}^{2}}\left( \overrightarrow{r}_{j}\times \overrightarrow{r}_{i}+%
\overrightarrow{r}_{k}\times \overrightarrow{r}_{j}+\overrightarrow{r}%
_{i}\times \overrightarrow{r}_{k}\right) \widehat{e}_{z}  \label{two3}
\end{equation}%
Introducing the following coordinate transformation%
\begin{gather}
\overrightarrow{r}_{ji}=\overrightarrow{r}_{j}-\overrightarrow{r}_{i}
\label{two3.1} \\
\overrightarrow{r}_{kj}=\overrightarrow{r}_{k}-\overrightarrow{r}_{j}  \notag
\\
\overrightarrow{R}=\overrightarrow{r}_{i}+\overrightarrow{r}_{j}+%
\overrightarrow{r}_{k}  \notag
\end{gather}%
eq.(\ref{two0}) reads%
\begin{equation}
U_{n_{i},n_{j},n_{k}}=\frac{1}{3}\int g_{n_{i}}(\left\vert \overrightarrow{r}%
_{ji}\right\vert ^{2})g_{n_{j}}(\left\vert \overrightarrow{r}%
_{kj}\right\vert ^{2})g_{n_{k}}(\left\vert \overrightarrow{r}_{ji}+%
\overrightarrow{r}_{kj}\right\vert ^{2})\cos (\frac{1}{6l_{B}^{2}}\left( 
\overrightarrow{r}_{kj}\times \overrightarrow{r}_{ji}\right) \widehat{e}%
_{z})d^{2}r_{ji}d^{2}r_{kj}d^{2}R  \label{two3.2}
\end{equation}%
finally we can introduce polar coordinates%
\begin{eqnarray}
x_{ji} &=&r_{ji}\cos \theta _{ji}\text{ \ \ \ \ \ \ \ \ \ }y_{ji}=r_{ji}\sin
\theta _{ji}  \label{two3.3} \\
\text{\ }x_{kj} &=&r_{kj}\cos \theta _{kj}\text{ \ \ \ \ \ \ \ \ }%
y_{kj}=r_{kj}\sin \theta _{kj}  \notag
\end{eqnarray}%
then, eq.(\ref{two3.2}) reads%
\begin{equation}
U_{n_{i},n_{j},n_{k}}=\sum\limits_{r=0}^{1}\sum\limits_{s=0}^{1}\sum%
\limits_{t=0}^{1}u_{n_{i}-r,n_{j}-s,n_{k}-t}(1-\delta
_{n_{i},0})^{r}(1-\delta _{n_{j},0})^{s}(1-\delta _{n_{k},0})^{t}
\label{two3.4}
\end{equation}%
where%
\begin{equation}
u_{a,b,c}=\frac{L^{2}}{3}\int
L_{a}(r_{ji}^{2})L_{b}(r_{kj}^{2})L_{c}((r_{ji}+r_{kj})^{2})\cos (\frac{1}{%
6l_{B}^{2}}r_{kj}r_{ji}\sin (\theta _{ji}-\theta _{kj}))r_{ji}dr_{ji}d\theta
_{ji}r_{kj}dr_{kj}d\theta _{kj}  \label{two3.4.0}
\end{equation}%
introducing the coordinates $s_{ji}=r_{ji}^{2}/2l_{B}^{2}$, $%
s_{kj}=r_{kj}^{2}/2l_{B}^{2}$, $\gamma =\theta _{ji}-\theta _{kj}$ and $\nu
=\theta _{ji}+\theta _{kj}$ and integrating over this last variable we obtain%
\begin{equation}
u_{a,b,c}=\frac{\pi L^{2}l_{B}^{4}}{3}T_{a,b,c}  \label{two3.4.1}
\end{equation}%
where%
\begin{equation}
T_{a,b,c}=\int e^{-(\sqrt{s_{ji}}+\sqrt{s_{kj}}%
)^{2}}L_{a}(s_{ji})L_{b}(s_{kj})L_{c}((\sqrt{s_{ji}}+\sqrt{s_{kj}})^{2})\cos
(\frac{\sqrt{s_{ji}s_{kj}}}{3}\sin (\gamma ))ds_{ij}ds_{kj}d\gamma
\label{two3.4.1.1}
\end{equation}%
We can perform the $\gamma $ integration%
\begin{equation}
\int_{0}^{2\pi }\cos (\frac{\sqrt{s_{ji}s_{kj}}}{3}\sin (\gamma ))d\gamma
=2\pi J_{0}(\frac{\sqrt{s_{ji}s_{kj}}}{3})  \label{two3.4.1.1.1}
\end{equation}%
then%
\begin{equation}
T_{a,b,c}=2\pi \int e^{-(\sqrt{s_{ji}}+\sqrt{s_{kj}}%
)^{2}}L_{a}(s_{ji})L_{b}(s_{kj})L_{c}((\sqrt{s_{ji}}+\sqrt{s_{kj}}%
)^{2})J_{0}(\frac{\sqrt{s_{ji}s_{kj}}}{3})ds_{ij}ds_{kj}
\label{two3.4.1.1.2}
\end{equation}%
This last integral is not computed in the work because of their complexity.%
\footnote{%
It will be source of future work to deduce higher order contributions to the
partition function of an electron gas in graphene placed in a constant
magnetic field.} Taking into account all the terms of the sum in eq.(\ref%
{two0}), the two permutation contribution to the partition function reads%
\begin{gather}
\int \gamma
_{n_{1},...,n_{N}}^{(2)}(r_{1},...,r_{N})d^{2}r_{1}...d^{2}r_{N}=L^{2(N-3)}2^{N-3}%
\frac{\pi L^{2}l_{B}^{4}}{3}\times  \label{two3.4.2} \\
\sum\limits_{i=1}^{N}\sum\limits_{j=1+1}^{N}\sum\limits_{k=j+1}^{N}\sum%
\limits_{r=0}^{1}\sum\limits_{s=0}^{1}\sum%
\limits_{t=0}^{1}T_{n_{i}-r,n_{j}-s,n_{k}-t}(1-\delta
_{n_{i},0})^{r}(1-\delta _{n_{j},0})^{s}(1-\delta _{n_{k},0})^{t}  \notag
\end{gather}%
and the partition function with the two permutation contribution reads%
\begin{equation}
Z_{N}(\beta )=\left( \sum\limits_{n=0}^{+n_{\Delta }}\cosh (\beta \Omega 
\sqrt{n})\right) ^{N}\eta _{N}L^{2N}\left( 1+\frac{\pi }{2L^{2}}%
2^{N-2}l_{B}^{2}F_{1}(\beta ,\Omega ,N)+\frac{\pi 2^{N-3}l_{B}^{4}}{3L^{4}}%
F_{2}(\beta ,\Omega ,N)\right)  \label{two3.4.3}
\end{equation}%
where%
\begin{equation}
F_{2}(\beta ,\Omega ,N)=\frac{\sum\limits_{n_{1}=0}^{+n_{\Delta
}}...\sum\limits_{n_{N}=0}^{+n_{\Delta }}\overset{N}{\underset{j=1}{\prod }}%
\cosh (\beta \Omega \sqrt{n_{j}})\sum\limits_{i=1}^{N}\sum%
\limits_{j=1+1}^{N}\sum\limits_{k=j+1}^{N}\sum\limits_{r=0}^{1}\sum%
\limits_{s=0}^{1}\sum\limits_{t=0}^{1}T_{n_{i}-r,n_{j}-s,n_{k}-t}}{\left(
\sum\limits_{n=0}^{+n_{\Delta }}\cosh (\beta \Omega \sqrt{n})\right) ^{N}}
\label{two3.4.4}
\end{equation}%
In this case, the sum has not been performed because we cannot solve eq.(\ref%
{two3.4.1.1.2}). The gauge dependent term $\sigma $ inhibit to consider the
statistical potential beyond one permutation and then has to be considered
an approximation, not only by the condition $\left\vert
r_{i}-r_{j}\right\vert >l_{B}$, but also for the correlation between three
or more electrons considered at once.

\section{Results and discussion}

We can use eq.(\ref{one10.1.1}) to compute thermodynamic functions. The
effective long wavelength approximation of the electron dynamics in graphene
in a uniform magnetic field is valid only at low energies, i.e. for
low-lying Landau levels as it was shown in the introduction. The properties
of higher Landau levels can be described within the tight-binding model by
introducing the Peierls substitution (see \cite{BA} and \cite{JH}). In
particular, we can consider an intense magnetic field in such a way that the
cutoff $n_{\Delta }$ is $n_{\Delta }\sim 1000$. This implies that $B\sim 
\frac{4\pi \hbar }{10^{3}3\sqrt{3}a^{2}e}\sim 100T$ which are magnetic
fields that can be obtained in laboratory. In particular, we will consider
only two particles, $N=2$, in this way, the correction to the partition
function of eq.(\ref{one10.1.1}) gives the total partition function of the
system.

\begin{figure}[tbp]
\centering
\includegraphics[width=105mm,height=70mm]{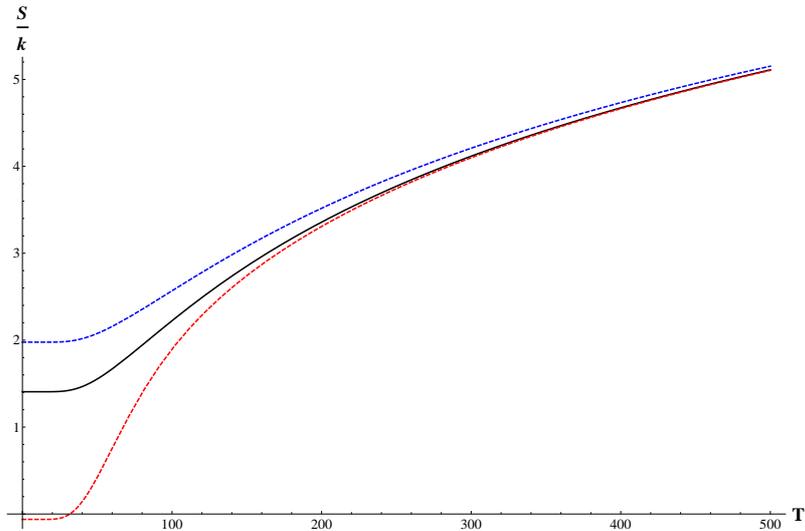}
\caption{Entropy as a function of the temperature. Without exchange
correlation (red line) and with exchange correlation (blue line). The black
line shows entropy without permutation correction.}
\label{entropy}
\end{figure}

\begin{figure}[tbp]
\centering
\includegraphics[width=105mm,height=70mm]{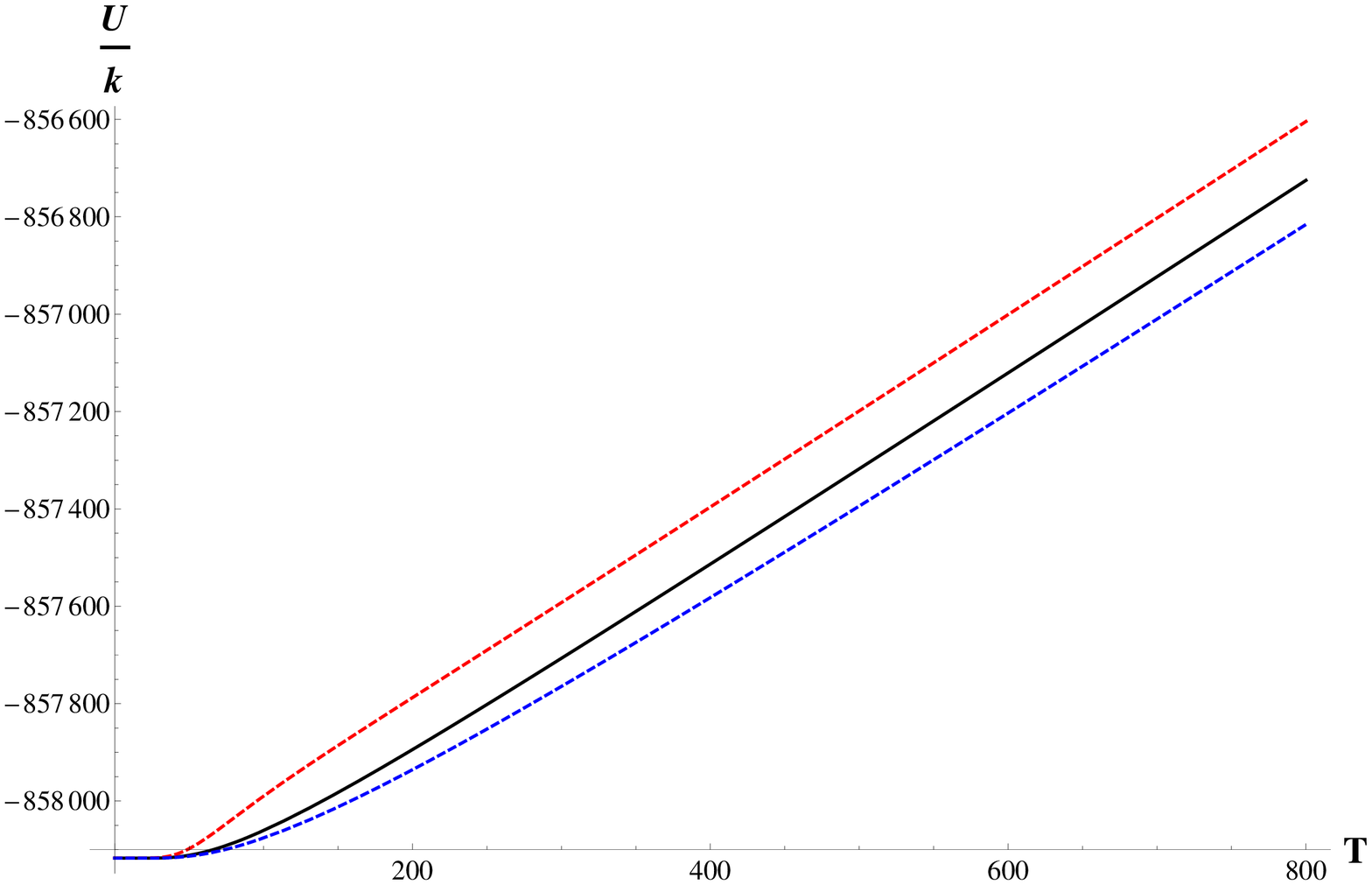}
\caption{Internal energy as a function of the temperature. Without exchange
correlation (red line) and with exchange correlation (blue line). The black
line shows entropy without permutation correction.}
\label{energy}
\end{figure}

\begin{figure}[tbp]
\centering
\includegraphics[width=105mm,height=70mm]{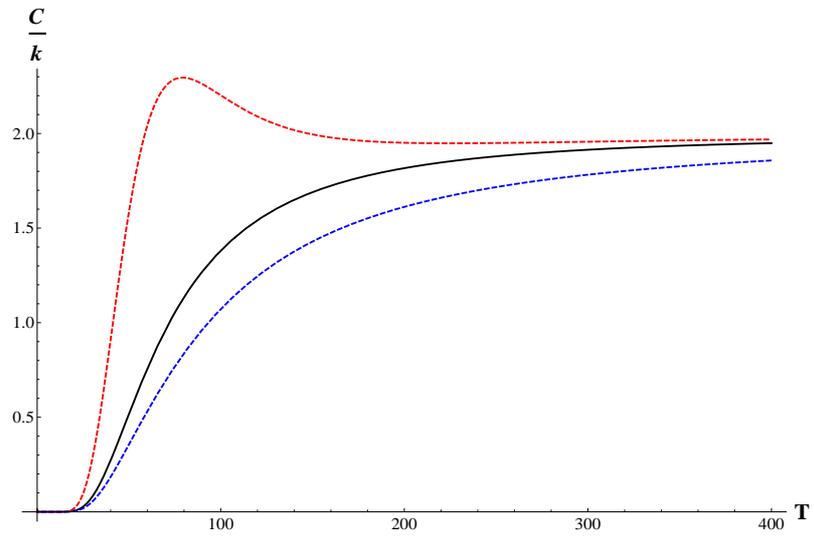}
\caption{Capacity heat as a function of the temperature. Without exchange
correlation (red line) and with exchange correlation (blue line). The black
line shows entropy without permutation correction.}
\label{specific heat}
\end{figure}

\begin{figure}[tbp]
\centering
\includegraphics[width=105mm,height=70mm]{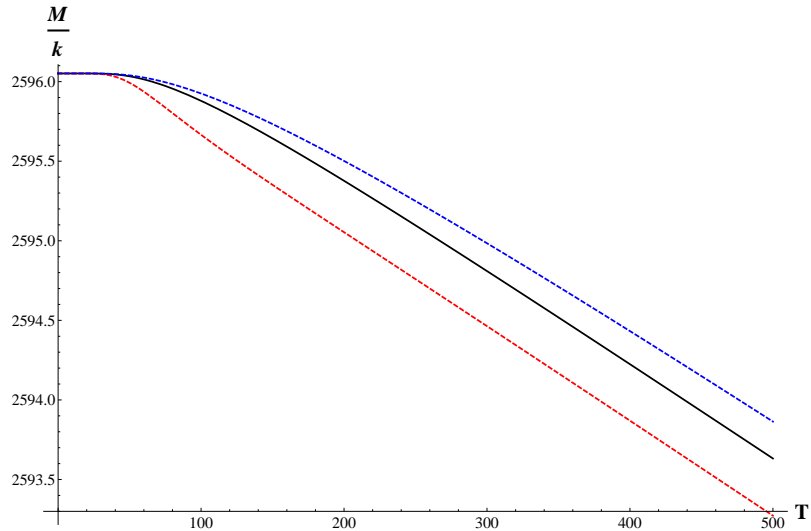}
\caption{Magnetic susceptibility as a function of temperature.Without
exchange correlation (red line) and with exchange correlation (blue line).
The black line shows entropy without permutation correction.}
\label{magnetic}
\end{figure}

Using the definition of the thermodynamic functions of eq.(\ref{bla1.3.9.2})
we can obtain the entropy, the internal energy, the specific heat and
magnetization for a quantum system composed of two Bloch electrons in a high
magnetic field for symmetric and antisymmetric states in the coordinates. In
figure \ref{entropy}, \ref{energy}, \ref{specific heat} and \ref{magnetic}
it is shown the results obtained with the partition function $%
Z_{2}^{(0)}(\beta )$ (black line in figures) and the partition function $%
Z_{2}^{(1)}(\beta )$ with the one permutation correction (red dashed line
for antisymmetric states and blue dashed line for symmetric states in
figures). In all the cases, the thermodynamic functions are plotted against
temperature in the range from $0$ to $1000$ K. 

As it can be seen in figure \ref{entropy}, the entropy of the quantum system
without interaction between Bloch electrons is larger than the entropy of
the same system in an anstisymmetric state and lower than the entropy of the
same system in a symmetric state in the coordinates. At higher temperatures,
both entropies match, which implies that disorder erase the exchange
correlation between electrons. At low temperatures, the difference between
entropies between antisymmetric states and symmetric states can be
understood due to the simple fact that there are more symmetric available
states than antisymmetric. The lowest energy value is $-2\Omega \sqrt{%
n_{\Delta }}$, which cannot support an antisymmetric state because it is
zero but it does support a symmetric state. As we move through the possible
energy levels of the composite system, more symmetric than antisymmetric
states are available. In fact, there are $(2n_{\Delta }+1)(2n_{\Delta }+2)/2$
symmetric states and $(2n_{\Delta }+1)n_{\Delta }$ antisymmetric states
available in the spectrum. In turn, the internal energy (see figure \ref%
{energy}) for antisymmetric states is larger than the internal energy of the
same system in a symmetric state. This is related to energy of the lowest
antisymmetric state available which is $E=-\Omega (\sqrt{n_{\Delta }}+\sqrt{%
n_{\Delta }+1})$.\footnote{%
We are not taking into account the Coulomb interaction, which can modify the
internal energy. It would be interesting to study the relative contribution
between Coulomb interaction and exchange correlation to the internal energy.}
The specific heat (see figure \ref{specific heat})\ of both statististics
has a limit behavior as $T\rightarrow \infty $, which correspond to entropy
saturation due to the upper bound in the energy levels. But the specific
heat of antisymmetric states has a peak, the Schottky anomaly, which is
related to the larger separation between consecutive energy values available
(see \cite{tari}). In \cite{hamze}, the specific heat is computed with doped
graphene in a magnetic field. In this case, the specific heat shows a
Schottky anomaly, but this is related to the band gap between conduction and
valence band introduced by the impurities (see \cite{TA}). 

The magnetization (see Figure \ref{magnetic}) shows a typical behavior of a
magnetic system, but in this case, the degrees of freedom that make the role
of spins are the conduction and valence bands. From the figure, the magnetic
susceptibility is negative and correspond to the ferromagnetic phase of
graphene in a magnetic field.

Finally, using the Helmholtz free energy, we can obtain the pressure of the
thermodynamic system by the following equation $P=-\frac{\partial F}{%
\partial A}$, where $A=L^{2}$.\ The general result reads%
\begin{equation}
P=kT\left[ \frac{N}{2A}-\frac{\frac{\pi }{8}l_{B}^{2}F_{1}(\frac{1}{kT}%
,\Omega ,N)}{A^{2}+\frac{\pi }{8}Al_{B}^{2}F_{1}(\frac{1}{kT},\Omega ,N)}%
\right]  \label{ga1}
\end{equation}

\begin{figure}[tbp]
\centering
\includegraphics[width=105mm,height=70mm]{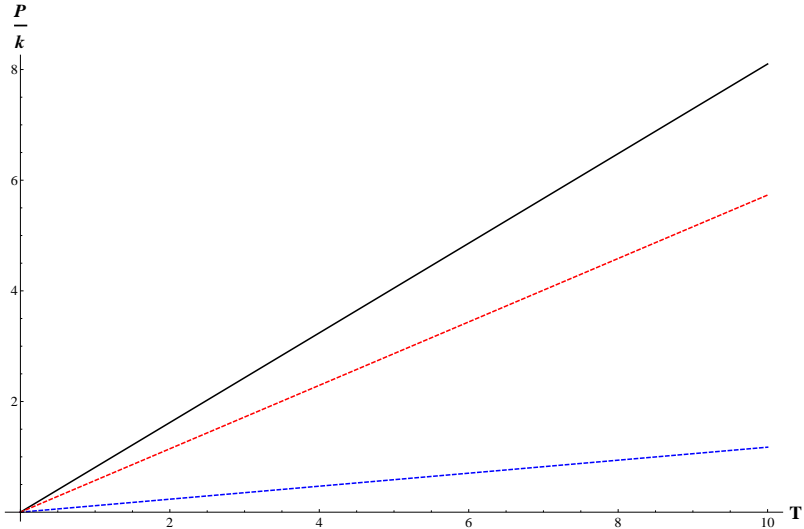}
\caption{Pressure as a function of temperature. The black line shows entropy
without permutation correction.}
\label{pressure}
\end{figure}

Using this last result for $N=3$, $\,$since with two particles, the
correction do not depend on $L^{2}$, the quantum pressure can be plotted
against temperature as it is shown in figure \ref{pressure}. 

As we can see in figure \ref{pressure}, the pressure tends to zero as $%
T\rightarrow 0$. A quantum degeneracy pressure would be expected, but the
degeneracy of fermions in graphene in a constant magnetic field introduce an
statistical potential that allows fermions to behave as attractive particles
at low temperature.

Due to the unusual thermal properties of electrons in graphene found in last
section when the exchange correlation term is considered in the partition
function, where an attractive statistical potential can be obtained, it
would be interesting to study the interaction between graphene-semiconductor
junction in the depletion region to see if there is any particular behavior
of the Schottky barrier. Theoretical and experimental works has been done in
this line of work (see \cite{1}, \cite{2}, \cite{3} and \cite{4}). 

\section{General considerations}

We can infer a general result for the succesive terms of the perturbation
expansion of the partition function that depends on the factor $\lambda =%
\frac{l_{B}}{L}\,$. The partition function can be written in a general form
as%
\begin{equation}
Z_{N}(\beta )=\sum\limits_{n_{1}=0}^{+n_{\Delta
}}...\sum\limits_{n_{N}=0}^{+n_{\Delta }}A_{n_{1},...,n_{N}}(N,\beta ,\Omega
)\eta _{N}L^{2N}2^{N}\sum\limits_{p=0}^{N-1}2^{-p}\lambda ^{p}C_{p}(N)
\label{gc2.0}
\end{equation}%
A factor $L^{2(N-p)}2^{N-p}$ comes from the $g_{n}$ function integration
that contains no permutation in its arguments. A $L^{2}l_{B}^{2p}$ factor
comes from the change of variable $s=r^{2}/2l_{B}^{2}$, where $L^{2}$ is due
to the center of mass variable integration. The coefficients $C_{p}(N)$
contains the results given by the integral of the $g_{n}$ functions with the
permuted arguments and for the first three coefficients we have obtained%
\begin{gather}
C_{0}(N)=1\text{, \ \ \ \ }C_{1}(N)=2\pi
\sum\limits_{i=1}^{N}\sum\limits_{j=i+1}^{N}\sum\limits_{l=0}^{1}\sum%
\limits_{k=0}^{1}\delta _{n_{i}-l,n_{j}-k}\text{\ }(1-\delta
_{n_{i},0})^{l}(1-\delta _{n_{j},0})^{k}  \label{gc2.1} \\
C_{2}(N)=2\pi
\sum\limits_{i=1}^{N}\sum\limits_{j=1+1}^{N}\sum\limits_{k=j+1}^{N}\sum%
\limits_{r=0}^{1}\sum\limits_{s=0}^{1}\sum%
\limits_{t=0}^{1}T_{n_{i}-r,n_{j}-s,n_{k}-t}(1-\delta
_{n_{i},0})^{r}(1-\delta _{n_{j},0})^{s}(1-\delta _{n_{k},0})^{t}  \notag
\end{gather}%
where $T_{n_{i}-r,n_{j}-s,n_{k}-t}$ is defined in eq.(\ref{two3.4.1.1.2}).
The factor $\eta _{N}L^{2N}$ in eq.(\ref{gc2.0}) contains the constant $%
\lambda ^{-N}$ which is common for all the terms.

The gauge dependent term $\sigma $ prevent us to understand the statistical
potential beyond one permutation. In turn, we cannot apply a cluster
expansion, where we separate particles that are close together from another
cluster because the cosine function in eq.(\ref{bla3.1.4}) contains the skew
product between position vectors. The unique possible way in which the
argument of the cosine function is zero is only when the particles are
aligned, but this is a very specific configuration of particles in graphene
sheet. In the case of two permutations, the vanishing of the cosine function
is translated to the vanishing of the Bessel function $J_{0}$, where the
argument is $r_{ji}r_{kj}/6l_{B}^{2}$.\ 

\section{Conclusion}

In this paper we have studied the thermodynamics properties of
non-interacting Bloch electrons in graphene in a constant magnetic field
computing the internal energy, specific heat, entropy and magnetic
susceptibility at zero and one order of the constant $l_{B}/L$ in the
perturbation expansion of the partition function and we have sketch the term
that contributes to second order. We have shown that at first order, the
partition function can be written as a system of $N$ Bloch electrons with a
attractive statistical potential for certain values of the interparticle
distance. Thermodynamic functions has been computed and compared between no
permutation and one permutation contribution for antisymmetric and symmetric
states. Due to the larger microstates available for symmetric states,
entropy increase its value with respect to same quantum system without one
permutation correction. Internal energy is lower for antisymmetric states
showing the effect of exchange correlation and the specific heat present a
Schottky anomaly due to the large gaps between energy levels available for
antisymmetric states. Graphene in a magnetic field present a magnetic
behavior, although no spin-orbit interaction has been taken into account.
The conducion and valence band acts as spin in graphene and couples to the
magnetic field with $\sqrt{B}$ dependence.

\section{Acknowledgment}

This paper was partially supported by grants of CONICET (Argentina National
Research Council) and Universidad Nacional del Sur (UNS) and by ANPCyT
through PICT 1770, and PIP-CONICET Nos. 114-200901-00272 and
114-200901-00068 research grants, as well as by SGCyT-UNS., E.A.G. and
P.V.J. are members of CONICET. P.B. and J. S.A. are fellow researchers at
this institution. 

\section{Appendix}

Suppose that our system consist of $N$ non-interacting Bloch electrons in
graphene in the long wavelength approximation, placed in a constant magnetic
field perpendicular to the graphene sheet. The eigenfunctions of the
Hamiltonian are those of eq.(\ref{intro2}) where the energies are those of
eq.(\ref{intro5}).

The partition function of the system reads%
\begin{equation}
Z_{N}(\beta )=Tr(e^{-\beta H})=\int \left\langle r_{1},..,r_{N}\left\vert
e^{-\beta H}\right\vert r_{1},..,r_{N}\right\rangle d^{2}r_{1}...d^{2}r_{N}
\label{ap1}
\end{equation}%
The eigenfunctions contain three indices, the quantum harmonic oscillator
index $n=0,1,2,...,n_{\Delta }$, the conduction and valence band $s=\pm 1$
and the wave vector $k$ in the $x$ direction. With these eigenfunctions we
can define an identity operator for a quantum system%
\begin{equation}
I=\sum\limits_{s=-1}^{+1}\sum\limits_{n=0}^{+n_{\Delta }}\int dk\left\vert
n,s,k\right\rangle \left\langle n,s,k\right\vert  \label{ap2}
\end{equation}%
We can introduce the identity for each quantum system in the partition
function, then%
\begin{gather}
Z_{N}(\beta
)=\sum\limits_{s_{1}=-1}^{+1}...\sum\limits_{s_{N}=-1}^{+1}\sum%
\limits_{n_{1}=0}^{+n_{\Delta }}...\sum\limits_{n_{N}=0}^{+n_{\Delta }}\exp
(-\beta \sum\limits_{j=1}^{N}E_{n_{j},s_{j}})\times  \label{ap3} \\
\int dk_{1}...dk_{N}\Phi _{\alpha _{1},...,\alpha _{N}}^{\ast
}(r_{1},...,r_{N})\Phi _{\alpha _{1},...,\alpha
_{N}}(r_{1},...,r_{N})d^{2}r_{1}...d^{2}r_{N}  \notag
\end{gather}%
where $\alpha _{j}=\{n_{j},s_{j},k_{j}\}$ are the index collection for the
Bloch electrons in graphene and%
\begin{equation}
\Phi _{\alpha _{1},...,\alpha _{N}}(r_{1},...,r_{N})=\frac{1}{\sqrt{N!}}%
\sum\limits_{P}^{{}}\delta _{P}P\{\psi _{\alpha _{1}}(r_{1})...\psi _{\alpha
_{N}}(r_{N})\}  \label{ap4}
\end{equation}%
where $\delta _{P}$ is $+1$ or $-1$ according as the permutation $P$ of the
single-particle wave functions $\psi _{\alpha _{j}}(r_{j})$ defined on eq.(%
\ref{intro2}) is even or odd, that is $\delta _{P}=(-1)^{[P]}$, where $[P]$
denotes the order of the permutation.\footnote{%
In this case we are considering only a antisymmetric quantum state in the
coordinates and a symmetric state in the spin variables. In the case we
consider a symmetric state in the coordinates, the permutation will contains
a plus sign.} The factor $(N!)^{-1/2}$ is introduced to secure the
normalization of the total wave function. Introducing eq.(\ref{ap4}) in eq.(%
\ref{ap3}) we have%
\begin{equation}
Z_{N}(\beta
)=\sum\limits_{s_{1}=-1}^{+1}...\sum\limits_{s_{N}=-1}^{+1}\sum%
\limits_{n_{1}=0}^{+n_{\Delta }}...\sum\limits_{n_{N}=0}^{+n_{\Delta }}\exp
(-\beta \sum\limits_{j=1}^{N}E_{n_{j},s_{j}})\times \int \chi _{\alpha
_{1},...,\alpha _{N}}(r_{1},...,r_{N})d^{2}r_{1}...d^{2}r_{N}  \label{ap5}
\end{equation}%
where 
\begin{equation}
\chi _{\alpha _{1},...,\alpha _{N}}(r_{1},...,r_{N})=\frac{1}{N!}\int
dk_{1}...dk_{N}\sum\limits_{P}^{{}}\sum\limits_{P^{\prime }}^{{}}\delta
_{P}\delta _{P^{\prime }}[\psi _{\alpha _{1}}^{\ast }(Pr_{1})\psi _{\alpha
_{1}}(P^{\prime }r_{1})...\psi _{\alpha _{N}}^{\ast }(Pr_{N})\psi _{\alpha
_{N}}(P^{\prime }r_{N})]  \label{ap6}
\end{equation}%
Taking outside the sum in $P$ and $P^{\prime }$ in last equation%
\begin{equation}
\chi _{\alpha _{1},...,\alpha _{N}}(r_{1},...,r_{N})=\frac{1}{N!}%
\sum\limits_{P}^{{}}\sum\limits_{P^{\prime }}^{{}}\delta _{P}\delta
_{P^{\prime }}\overset{N}{\underset{j=1}{\prod }}f_{\alpha
_{j}}(Pr_{j},P^{\prime }r_{j})  \label{ap7}
\end{equation}%
where%
\begin{equation}
f_{\alpha _{j}}(Pr_{j},P^{\prime }r_{j})=\int dk_{j}\psi _{\alpha
_{j}}^{\ast }(Pr_{j})\psi _{\alpha _{j}}(P^{\prime }r_{j})  \label{ap8}
\end{equation}%
is a function of the coordinates and can be evaluated by replacing eq.(\ref%
{intro2}) into last equation%
\begin{equation}
f_{\alpha _{j}}(Pr_{j},P^{\prime }r_{j})=\frac{\left\vert
C_{n_{j}}\right\vert ^{2}}{2L}\int dk_{j}e^{ik_{j}(P^{\prime
}x_{j}-Px_{j})}\varphi _{(n_{j},s_{j},k_{j})}^{T}(P\xi _{j})\varphi
_{(n_{j},s_{j},k_{j})}(P^{\prime }\xi _{j})  \label{ap9}
\end{equation}%
where $\varphi _{(n_{j},s_{j},k_{j})}(\xi _{j})$ is defined on eq.(\ref%
{intro3}). Introducing eq.(\ref{intro3}) on eq.(\ref{ap9}) we obtain%
\begin{gather}
f_{\alpha _{j}}(Pr_{j},P^{\prime }r_{j})=\frac{1}{2(2-\delta _{n_{j},0})L}%
\times  \label{ap10} \\
\int dk_{j}e^{ik_{j}(P^{\prime }x_{j}-Px_{j})}(2s_{j}^{2}\phi
_{n_{j}-1,k_{j}}^{\ast }(P\xi _{j})\phi _{n_{j}-1,k_{j}}(P^{\prime }\xi
_{j})(1-\delta _{n_{j},0})+2\phi _{n_{j},k_{j}}^{\ast }(P\xi _{j})\phi
_{n_{j},k_{j}}(P^{\prime }\xi _{j}))  \notag
\end{gather}%
and using eq.(\ref{intro4})%
\begin{gather}
f_{\alpha _{j}}(Pr_{j},P^{\prime }r_{j})=\frac{\pi ^{-1/2}}{(2-\delta
_{n_{j},0})L}\int dk_{j}e^{ik_{j}(P^{\prime }x_{j}-Px_{j})}\times
\label{ap11} \\
(\frac{s_{j}^{2}}{2^{n_{j}-1}(n_{j}-1)!}e^{-\frac{1}{2}((P\xi
_{j})^{2}+(P^{\prime }\xi _{j})^{2})}H_{n_{j-1,k_{j}}}(P\xi
_{j})H_{n_{j-1,k_{j}}}(P^{\prime }\xi _{j})(1-\delta _{n_{j},0})+  \notag \\
\frac{1}{2^{n_{j}}n_{j}!}e^{-\frac{1}{2}((P\xi _{j})^{2}+(P^{\prime }\xi
_{j})^{2})}H_{n_{j,k_{j}}}(P\xi _{j})H_{n_{j,k_{j}}}(P^{\prime }\xi _{j})) 
\notag
\end{gather}%
Completing squares in the exponentials we obtain for the first integral in
last equation\footnote{%
For the second integral the same result holds with the $n_{j}-1\rightarrow
n_{j}$ replacement.}%
\begin{gather}
f_{\alpha _{j}}^{(1)}(Pr_{j},P^{\prime }r_{j})=\frac{s_{j}^{2}\pi
^{-1/2}e^{b}}{(2-\delta _{n_{j},0})L2^{n_{j}-1}(n_{j}-1)!}\times
\label{ap12} \\
\int dk_{j}e^{-(l_{B}k_{j}-l_{B}a)^{2}}H_{n_{j-1,k_{j}}}(\frac{Py_{j}}{l_{B}}%
-l_{B}k_{j})H_{n_{j-1,k_{j}}}(\frac{y_{j}}{l_{B}}-l_{B}k_{j})  \notag
\end{gather}%
where the supperscript in $f_{\alpha _{j}}^{(1)}(Pr_{j},P^{\prime }r_{j})$
indicate the first integral of eq.(\ref{ap11}) and%
\begin{equation}
b(Px_{j},P^{\prime }x_{j},Py_{j},P^{\prime }y_{j})=\frac{(Py_{j}+P^{\prime
}y_{j}+i(P^{\prime }x_{j}-Px_{j}))^{2}}{4l_{B}^{2}}-\frac{1}{2l_{B}^{2}}%
((Py_{j})^{2}+(P^{\prime }y_{j})^{2})  \label{ap13}
\end{equation}%
and 
\begin{equation}
a(Px_{j},P^{\prime }x_{j},Py_{j},P^{\prime }y_{j})=\frac{(Py_{j}+P^{\prime
}y_{j}+i(P^{\prime }x_{j}-Px_{j}))}{2l_{B}^{2}}  \label{ap14}
\end{equation}%
By introducing the following coordinate transformation 
\begin{equation}
q_{j}=\frac{Py_{j}}{l_{B}}-l_{B}k_{j}  \label{ap15}
\end{equation}%
eq.(\ref{ap12}) reads%
\begin{gather}
f_{\alpha _{j}}^{(1)}(Pr_{j},P^{\prime }r_{j})=-\frac{s_{j}^{2}\pi
^{-1/2}e^{b}}{(2-\delta _{n_{j},0})Ll_{B}2^{n_{j}-1}(n_{j}-1)!}\times
\label{ap16} \\
\int dq_{j}e^{-(\frac{Py_{j}}{l_{B}}%
-q_{j}-l_{B}a)^{2}}H_{n_{j-1,k_{j}}}(q_{j})H_{n_{j-1,k_{j}}}(\frac{P^{\prime
}y_{j}}{l_{B}}+q_{j}-\frac{Py_{j}}{l_{B}})  \notag
\end{gather}%
A second coordinate transformation can be applied 
\begin{equation}
-z_{j}=\frac{Py_{j}}{l_{B}}-q_{j}-l_{B}a  \label{ap17}
\end{equation}%
that change eq.(\ref{ap16}) in%
\begin{gather}
f_{\alpha _{j}}^{(1)}(Pr_{j},P^{\prime }r_{j})=-\frac{s_{j}^{2}\pi
^{-1/2}e^{b}}{(2-\delta _{n_{j},0})Ll_{B}2^{n_{j}-1}(n_{j}-1)!}\times
\label{ap18} \\
\int dz_{j}e^{-z_{j}^{2}}H_{n_{j-1,k_{j}}}(\frac{Py_{j}}{l_{B}}%
-l_{B}a+z_{j})H_{n_{j-1,k_{j}}}(\frac{y_{j}}{l_{B}}+z_{j}-l_{B}a)  \notag
\end{gather}%
Using eq.(7.377$^{8}$) of page 804 of \cite{laguerre}%
\begin{equation}
\int e^{-x^{2}}H_{m}(x+y)H_{n}(x+z)dx=2^{n}\pi
^{1/2}m!z^{n-m}L_{m}^{n-m}(-2yz)  \label{ap19}
\end{equation}%
where $m\leq n$ and $L_{m}^{n-m}(x)$ is the associated Laguerre polynomial,
eq.(\ref{ap18}) reads 
\begin{equation}
f_{\alpha _{j}}^{(1)}(Pr_{j},P^{\prime }r_{j})=-\frac{s_{j}^{2}e^{b}}{%
(2-\delta _{n_{j},0})Ll_{B}}L_{n_{j}-1}(-2(\frac{Py_{j}}{l_{B}}-l_{B}a)(%
\frac{P^{\prime }y_{j}}{l_{B}}-l_{b}a))  \label{ap20}
\end{equation}%
Finally, taking into account that the same result is obtained in the second
integral in eq.(\ref{ap11}) with the $n_{j}-1\rightarrow n_{j}$ replacement,
eq.(\ref{ap18}) reads 
\begin{gather}
f_{\alpha _{j}}(Pr_{j},P^{\prime }r_{j})=-\frac{e^{b}}{(2-\delta
_{n_{j},0})Ll_{B}}\times  \label{ap21} \\
\left( L_{n_{j}-1}(-2(\frac{Py_{j}}{l_{B}}-l_{B}a)(\frac{P^{\prime }y_{j}}{%
l_{B}}-l_{b}a))(1-\delta _{n_{j},0})+L_{n_{j}}(-2(\frac{Py_{j}}{l_{B}}%
-l_{B}a)(\frac{P^{\prime }y_{j}}{l_{B}}-l_{b}a))\right)  \notag
\end{gather}%
by doins some algebraic manipulations we can rewrite the argument of the
Laguerre polynomials as%
\begin{equation}
-2(\frac{Py_{j}}{l_{B}}-l_{B}a)(\frac{P^{\prime }y_{j}}{l_{B}}-l_{B}a)=\frac{%
1}{2l_{B}^{2}}\left\vert P^{\prime }r_{j}-Pr_{j}\right\vert ^{2}
\label{ap22}
\end{equation}%
where $Pr_{j}=(Px_{j},Py_{j})$. In turn, the $b$ function can be written as%
\begin{equation}
b=-\frac{1}{4l_{B}^{2}}(\left\vert Pr_{j}-P^{\prime }r_{j}\right\vert
^{2})+i\sigma  \label{ap23}
\end{equation}%
where $\sigma $ is the gauge-dependent term which reads\footnote{%
These gauge term can be found in the Green function of the same quantum
system (see \cite{dyna2}, eq.(3)).}%
\begin{equation}
\sigma (Pr_{j},P^{\prime }r_{j})=\frac{1}{2l_{B}^{2}}((P^{\prime
}y_{j}+Py_{j})(P^{\prime }x_{j}-Px_{j}))  \label{ap24}
\end{equation}%
Using these results, the function $f_{\alpha _{j}}(Pr_{j},P^{\prime }r_{j})$
finally reads%
\begin{gather}
f_{\alpha _{j}}(Pr_{j},P^{\prime }r_{j})=-\frac{e^{-\frac{1}{4l_{B}^{2}}%
\left\vert Pr_{j}-P^{\prime }r_{j}\right\vert ^{2}+i\sigma }}{(2-\delta
_{n_{j},0})Ll_{B}}\times  \label{ap25} \\
\left[ L_{n_{j}-1}(\frac{1}{2l_{B}^{2}}\left\vert Pr_{j}-P^{\prime
}r_{j}\right\vert ^{2})(1-\delta _{n_{j},0})+L_{n_{j}}(\frac{1}{2l_{B}^{2}}%
\left\vert Pr_{j}-P^{\prime }r_{j}\right\vert ^{2})\right]  \notag
\end{gather}%
This result will be used in the main sections of this paper.

\bigskip 


\begin{thebibliography}{99}
\bibitem{novo} K. S. Novoselov, A. K. Geim, S. V. Morozov, D. Jiang, M. I.
Katsnelson, I. V. Grigorieva, S. V. Dubonos and A. A. Firsov, \textit{Nature}%
, \textbf{438}, 197 (2005).

\bibitem{intro1} A.K. Geim and K. S. Novoselov, \textit{Nature Materials,} 
\textbf{6}, 183 (2007).

\bibitem{intro2} Y. B. Zhang, Y.W. Tan, H. L. Stormer and P. Kim, \textit{%
Nature}, \textbf{438}, 201 (2005).

\bibitem{B} A. H. Castro Neto, F. Guinea, N. M. R. Peres, K. S. Novoselov
and A. K. Geim, \textit{Rev. Mod. Phys.},\textbf{\ 81}, 109 (2009).

\bibitem{BBBB} M. O. Goerbig, \textit{Rev. Mod. Phys.}, \textbf{83}, 4,
(2011).

\bibitem{A} J. McClure, \textit{Phys. Rev.,} \textbf{104}, 666 (1956).

\bibitem{kuru} S. Kuru, J. Negro and L. M. Nieto, \textit{J. Phys.: Condens.
Matter}, \textbf{21}, 455305 (2009).

\bibitem{E} Y. Zheng and T. Ando, \textit{Phys. Rev. B}, \textbf{65, }245420
(2002).

\bibitem{NA1} A. Matulis and F. M. Peters, \textit{Phys. Rev. B,} \textbf{75,%
} 125429 (2007).

\bibitem{NA2} R. Nasir, K. Sabeeh and M. Tahir, \textit{Phys. Rev. B,} 
\textbf{81}, 085402 (2010).

\bibitem{NA3} M. Tahir and K. Sabeeh, \textit{Phys. Rev. B,} \textbf{77,}
195421 (2008).

\bibitem{NA4} T.Z. Li, K. I. Wang and J. I. Wang, \textit{J. Phys.: Condens.
Matter,} \textbf{9} 9299 (1997).

\bibitem{NA6} A.R. Wright, J. Liu, Z. Ma, Z. Zeng, W. Xu and C. Zhang, 
\textit{Microelectronics Journal,} \textbf{40}, 716--718 (2009).

\bibitem{NA5} R. Nasir, M. A. Khan, M. Tahir and K. Sabeeh, \textit{J.
Phys.: Condens. Matter,} \textbf{22,} 025503 (2010).

\bibitem{hamze} H. Mousavi, \textit{Physica B, }\textbf{414}, 78--82 (2013).

\bibitem{Huang} K. Huang, Statistical mechanics, John Wiley \& Sons, New
York, (1987).

\bibitem{peres-guinea} N. M. Peres, F. Guinea and H. Castro Neto, \textit{%
Phys. Rev. B}, \textbf{73}, 125411 (2006).

\bibitem{dyna1} P. K. Pyatkovskiy and V. P. Gusynin, \textit{Phys. Rev. B}, 
\textbf{83}, 075422 (2011)

\bibitem{dyna2} T. M. Rusin\ and W. Zawadzki, \textit{J. Phys. A: Math.
Theor. }\textbf{44,} 105201 (2011).

\bibitem{laguerre} I. S. Gradshtein and I. M. Ryzhik, Table of Integrals,
Series, and Products, New York: Academic Press, page 804 (2007).

\bibitem{lambert} R. Corless, G. Gonnet, D. Hare, D. Jeffrey, D. Knuth, On
the Lambert W function, Advances in Computational Mathematics, Berlin, New
York: Springer-Verlag, 5: 329--359 (1996).

\bibitem{tapash} T. Chakraborty,V. M. Apalkov, \textit{Solid State Commun.},
http://dx.doi.org/10.1016/j.ssc.2013.04.002i, (2013).

\bibitem{nomura} K. Nomura and A.H. MacDonald, \textit{Phys. Rev. Lett.} 
\textbf{96, }p. 256602 (2006).

\bibitem{goerbig2} .M.O. Goerbig, R. Moessner and B. Doucot, \textit{Phys.
Rev. B} \textbf{74} (2006), p. 161407.

\bibitem{BA} B.A. Bernevig, T.L. Hughes, S.C. Zhang, H.D. Chen, and C. Wu, 
\textit{Int. J. Mod. Phys. B,} \textbf{20, }p. 3257 (2006).

\bibitem{JH} J.H. Ho, Y.H. Lai, Y.H. Chiu and M.F. Lin, \textit{Physica E}, 
\textbf{40,} pp. 1722--1725(2008).

\bibitem{tari} A. Tari, The Specific Heat of Matter at Low Temperatures,
Imperial College Press, (2003).

\bibitem{TA} T. Altanhana and B. Kozal, \textit{Eur. Phys. J. B}, \textbf{85}%
: 222 (2012).

\bibitem{1} J. Lua, B. Xub, H. Liua , Y. Wanga, W. Zhengc, \textit{%
Superlattices and Microstructures,} \textbf{60,} 217--223 (2013).

\bibitem{2} C. Chen, W. Zhang, B. Zhao, Y. Zhang, \ \textit{Phys. Lett. A}, 
\textbf{374,} 309--312 (2009).

\bibitem{3} L. Lancellotti, T. Polichetti, F. Ricciardella, O. Tari, S.
Gnanapragasam, S. Daliento, G. Di Francia, \textit{Thin Solid Films,} 
\textbf{522,} 390--394 (2012).

\bibitem{4} S. Tongay, T. Schumann, X. Miao, B.R. Appleton, A.F. Hebard, 
\textit{Carbon},  \textbf{49,} 2033--2038 (2011).
\end{thebibliography}
\end{document}